\documentclass[english,review]{revtex4-1}
\usepackage[utf8]{inputenc}
\usepackage{babel}
\usepackage{array}
\usepackage{float}
\usepackage{amsmath}
\usepackage{graphicx}
\usepackage{subscript}
\usepackage[unicode=true,
 bookmarks=false,
 breaklinks=false,pdfborder={0 0 1},backref=section,colorlinks=false]
 {hyperref}

\makeatletter

\providecommand{\tabularnewline}{\\}

\usepackage{lineno}
\usepackage{babel}
\usepackage{setspace}
\usepackage{subfloat}
\usepackage{xcolor}

\@ifundefined{showcaptionsetup}{}{%
 \PassOptionsToPackage{caption=false}{subfig}}
\usepackage{subfig}
\makeatother

\begin{document}

\title{Structure and dielectric properties of Ba$_{2}$Cu$_{x}$Y$_{1-x}$TaO$_{6-y}$
double perovskite}

\author{F. S. Oliveira\textsuperscript{1{*}} ,C. A. M. dos Santos\textsuperscript{1}
, A. J. S. Machado\textsuperscript{1}, P. Banerjee\textsuperscript{2 }and
A. Franco Jr.\textsuperscript{3}}

\address{\textsuperscript{1}Escola de Engenharia de Lorena, Universidade
de São Paulo, Lorena, Brazil}
\email{fso@usp.br}

\selectlanguage{english}%

\address{\textsuperscript{2}Department of Physics, Gandhi Institute of Technology
and Management (GITAM) University, Bengaluru, India}

\address{\textsuperscript{3}Instituto de Física, Universidade Federal de
Goiás, Goiania, Brazil}
\begin{abstract}
\begin{center} \textbf{Abstract} \end{center}
In this paper, we reported the effect of Cu doping on the structural
and dielectric properties of Ba$_{2}$Y$_{1-x}$Cu$_{x}$TaO$_{6-y}$
(0.00 $\leq$ x $\leq$ 0.50) ceramics at room temperature. The Copper
for Yttrium substitution reduces the sintering temperature and leads
to structural changes in the Ba$_{2}$YTaO$_{6}$ rock-salt crystalline
structure. Dielectric permittivity and complex impedance spectroscopy
measurements suggested enhancement of the dielectric constant and
occurrence of interfacial Maxwell-Wagner polarization. 
\end{abstract}
\maketitle

\section{Introduction}

Systematic analysis of YBa$_{2}$Cu$_{3}$O$_{7-\delta}$ (Y123) superconductor
shows that Ta$_{2}$O$_{5}$ helps to achieve large-grains
\citep{oliveira2005,bortolozo2004} for the magnetic levitation and magnets applications \citep{bondarenko2017,siems2004,yu1997}.
The grain growth is attributed to the decrease in the peritectic temperature
transformation \citep{bortolozo2004}, and the homogeneous segregation
of Ba$_{2}$YTaO$_{6}$ (BYT) secondary phase in the Y123 superconducting
matrix acts as vortex pinning centers,
which are non-superconducting regions that confine the quantum magnetic
flux, and consequently, optimizes the critical current density parameter
\citep{oliveira2005,coll2014}. Futhermore, BYT phase also induces an
unusual Paramagnetic Meissner Effect in Y123 superconductor \citep{dias2014}.

Single Crystals of BYT were grown by Galasso \textit{et al.} using
B$_{2}$O$_{3}$ flux in 1960's \citep{galasso1966}. Later the dielectric
properties of BYT ceramic were studied in the microwave frequency
range \citep{zurmuhlen1994}, and a structural phase transition \citep{gupta2013}
from cubic to tetragonal space group around 260\,K was observed by
diffractometry, calorimetry, transmission electronic microscopy, and
Raman scattering \citep{zurmuhlen1994,lufaso2006,zhou2011}.
This transition is quite similar to the SrTiO$_{3}$ antiferrodistortive
case \citep{de2015}.

BYT belongs to the family of complex perovskite \citep{king2010}
oxides. This family group has attracted a lot of attention due to
the presence of a large number of oxide materials which can be formed
\citep{vasala2015a2b}. The long-range order of the crystal lattice
are responsible for several properties in these complex perovskites
\citep{king2010,vasala2015a2b} such as the high dielectric permittivity
\citep{yanez2009,chandrasekhar2012}. On the other hand the extrinsic
giant dielectric permittivity frequency observed in related perovskites
such as CaCu$_{3}$Ti$_{4}$O$_{12}$ (CCTO) \citep{subramanian2000}
is a consequence of semiconducting grains limited by insulating grain
boundaries which act as a kind of barrier to the free carriers motion
inside the grains \citep{adams2002,tien1967,waser1991}. The dielectric relaxation at these barriers of charge is well described
by the Maxwell-Wagner (M-W) polarization model \citep{wang2013,o2000,banerjee2017}.

Since as BYT appears as a secondary phase in Ta-doped Y123 \citep{oliveira2005}, it suggests to one
that Copper should have some solubility in the BYT complex perovskite,
such as the Ba$_{2}$Y$_{1-x}$Cu$_{x}$WO$_{6-y}$ system already reported \citep{garcia1992}. This paper presents the first investigation of the physical properties of Ba$_{2}$Y$_{1-x}$Cu$_{x}$TaO$_{6-y}$ system for 0.00 $\leq$ x $\leq$ 0.50. Our results demonstrate that the sintering process, crystal structure, and dielectric relaxation change are dependent of the sample composition. We also show that Copper enhances the dielectric constant of BYT ceramics and leads to interfacial Maxwell-Wagner polarization at the grain boundaries. The extrinsic effects induced by Cu turns Ba$_{2}$Y$_{1-x}$Cu$_{x}$TaO$_{6-y}$ ceramics new candidates for some applications in electronic devices.

\section{Experimental Procedure}

Ceramics of Ba$_{2}$Y$_{1-x}$Cu$_{x}$TaO$_{6-y}$ (0.00 $\leq$
x $\leq$ 0.50) were prepared by the standard solid state reaction
method using BaCO$_{3}$ (Sigma-Aldrich, 99.95\%) Y$_{2}$O$_{3}$
(Sigma-Aldrich, 99.99\%) Ta$_{2}$O$_{5}$ (Cerac, 99.99\%) and CuO
(Sigma-Aldrich, 99.99\%) powders. For each stoichiometry, powders
were ball milled for about 12 hours, then calcined at 950$^{\circ}$\,C
in the air during 96 hours. Approximately 10\% in mass of polyvinyl
alcohol (PVA) binder additive was added to the powders, and disc pellets
with about 8 mm in diameter and 2 mm in thickness were pressed uniaxially.
These pellets were annealed at 500$^{\circ}$\,C for 1 hour to decompose
the organic PVA \citep{banerjee2016}, and thereafter sintered in
air at 1250$^{\circ}$\,C to 1400$^{\circ}$\,C for up to 120 hours. The microstructure of ceramics were analyzed by Scanning electron microscopy (SEM).

The crystalline structure of all samples was analyzed by XRD technique
using Empirian Pan Analytical diffractometer with CuK$\alpha$ radiation
($\lambda$ = 1.5406 {Å}) and Ni filter. The diffractometry measurements
were carried out with 0.01$^{\circ}$ step in $10^{\circ}\leq2\theta\leq90^{\circ}$
range. Model refinement profiles with Pseudo-Voigt function by Rietveld
method were performed in HiScore Plus program using information from
inorganic crystal structure database (ICSD) \citep{belsky2002}. Both
subdomain size and microstrain were obtained from Williamson-Hall
plots \citep{woodward1994,alves2017,dhahri2018}. The dielectric properties
of the ceramic samples were studied using a computer controlled Agilent
4980A LCR meter. An alternating voltage of 1.0\,V was applied on
the ceramic pellets with silver painted faces over 20\,Hz to 2\,MHz
frequency. Nyquist plots were analyzed using EIS Spectrum Analyzer
Program \citep{bondarenko2005inverse}.

\section{Results and Discussion}

\begin{figure}[H]
\centering \subfloat[]{\includegraphics[width=0.7\textwidth]{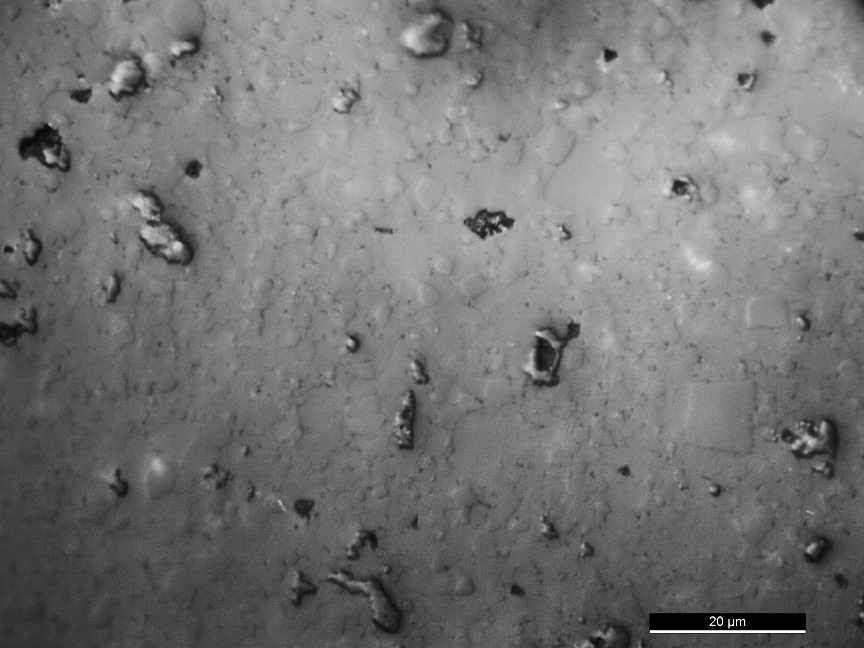}

}\\
 \subfloat[]{\includegraphics[width=0.7\textwidth]{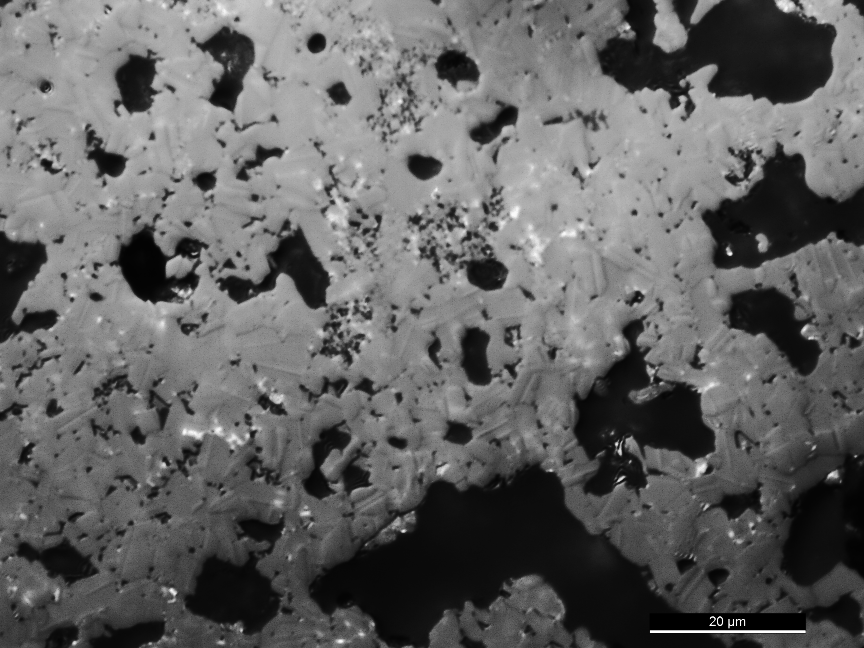}

}\caption{SEM images for (a) x = 0.00 and (b) x = 0.40 sample.}
\label{fig:SEM} 
\end{figure}

The pure Ba$_{2}$YTaO$_{6}$ was formed only after long sintering
time, 120 hours at 1400$^{\circ}$\,C while the Cu-doped Ba$_{2}$Y$_{1-x}$Cu$_{x}$TaO$_{6-y}$
(x=0.40) was formed after much shorter sintering time, 15 hours at
1250$^{\circ}$\,C. Thus it was evident that Cu for Y substitution
lower both sintering temperature and sintering
time. The figure \ref{fig:SEM} shows that x = 0.40 sample (fig. \ref{fig:SEM}(b)) has signatures of liquid phase and more defects when compared with x = 0.00 sample (fig. \ref{fig:SEM}(a)). It is an evidence that the Cu concentration contributes significantly to the microstructure.

\begin{figure}[H]
\centering \subfloat[]{\includegraphics[width=0.7\textwidth]{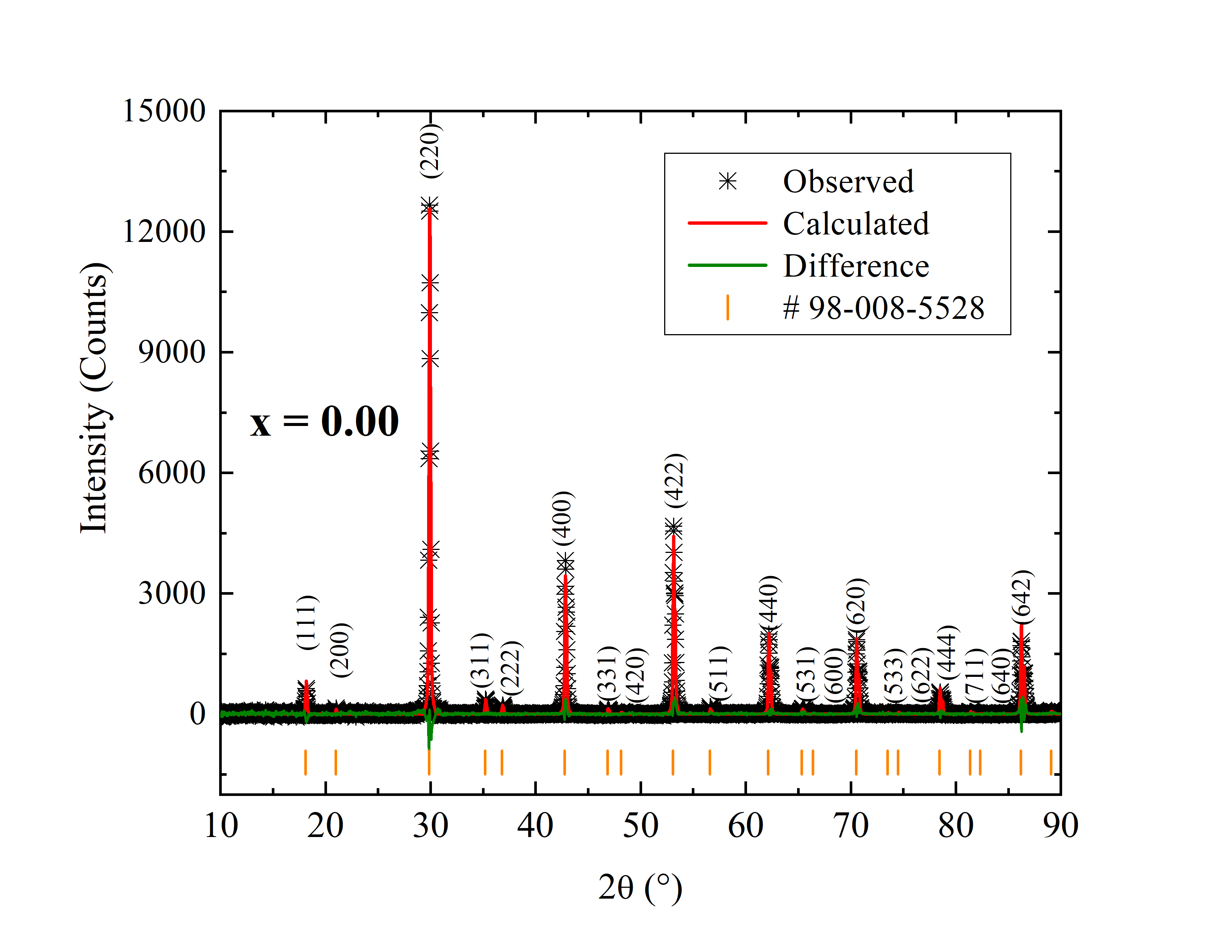}

}\\
 \subfloat[]{\includegraphics[width=0.7\textwidth]{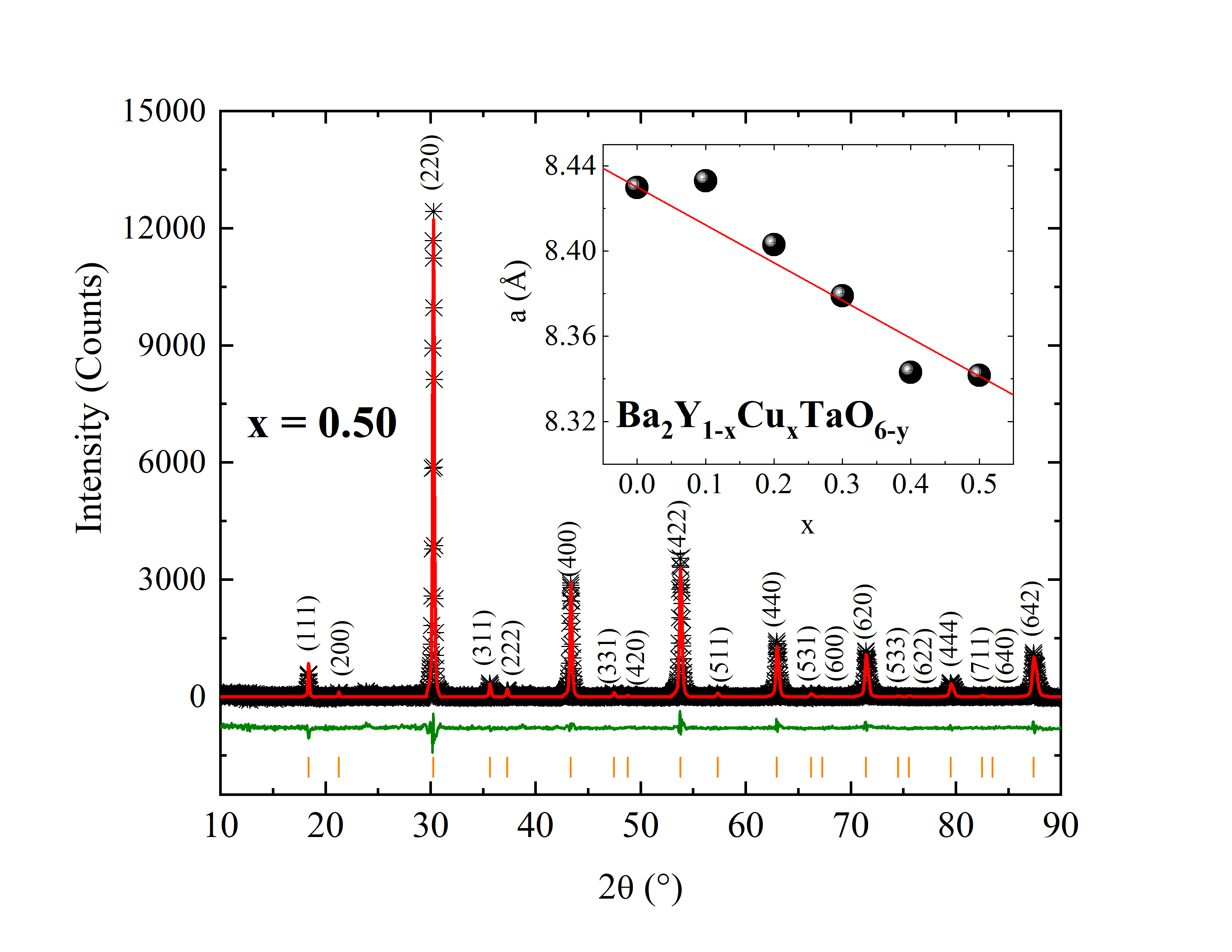}

}\caption{(a) XRD pattern of x = 0.00 sample sintered at 1450$^{\circ}$\,C
for 120 hours. (b) XRD pattern of x = 0.50 sample sintered at 1250$^{\circ}$\,C
for 15 hours. The inset shows the Vegard's law dependence between
lattice parameter and copper content.}
\label{fig:XRD} 
\end{figure}

Figure \ref{fig:XRD} exhibits the XRD patterns for x = 0.00 and x
= 0.50 samples. Our analysis suggest that the sample with the highest
Cu doping level possesses a single cubic perovskite crystallographic
phase. The lattice parameter of the undoped sample is in good agreement
with those reported in the literature \citep{galasso1966,zurmuhlen1994,lufaso2006,zhou2011}
and the decrease in the lattice parameter (see inset of the figure
\ref{fig:XRD}(b) can be attributed to the differences in the ionic
radii of Cu$^{2+}$ (0.73\,{Å}) and Y$^{3+}$ (0.9\,{Å}) ions
\citep{shannon1976}.

\begin{figure}[H]
\centering \includegraphics[width=0.7\textwidth]{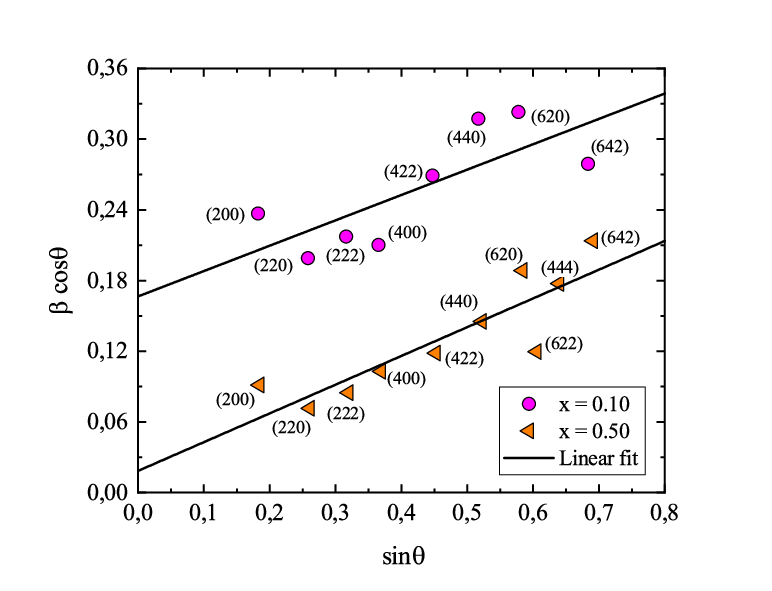}\\
 \caption{Williamson-Hall plot of x = 0.10 and x = 0.50, where $\beta$ is the
full width at half maximum of the reflection peaks.}
\label{fig:WH} 
\end{figure}

The Williamson-Hall plot shown in Figure \ref{fig:WH} gives the strain
and the sub-cell domain information from the slope and the reciprocal
of the intercept respectively \citep{woodward1994}. The Rietveld
refinement results obtained for all samples are shown in the Table
\ref{tab:rietveld}. The Cu for Y substitution produce compensating oxygen vacancies may induce strain in the crystal lattice. Additionally, XRD refinement also suggests that substitution increases the crystallite size in Ba$_{2}$Y$_{1-x}$Cu$_{x}$TaO$_{6-y}$ system.

\begin{figure}[H]
\centering \includegraphics[width=0.7\textwidth]{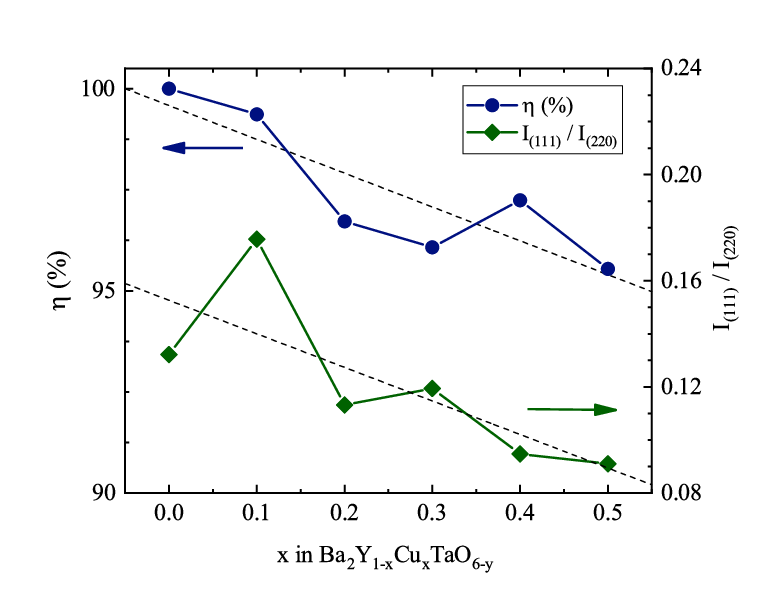}\\
 \caption{Ordering degree (left-axis) and ratio between (111) and (220) peak
intensities (right-axis). The dashed lines are just a guides to the
eyes.}
\label{fig:ordering} 
\end{figure}

In a double perovskite, the quantity of the long-range ordering degree
($\eta$) is given by \citep{liu2003}:

\begin{equation}
\eta=2\left|M_{0}-0.5\right|\label{eq:desordem}
\end{equation}

where $M_{0}$ is the refined occupation of Ta in the 4a (0,0,0)
site or the refined occupation of Y or Cu in the 4b
(1/2,1/2,1/2) site \citep{woodward1994}. Beyond that, superlattice
reflection peaks are sharper in high-ordered perovskites and the ordering
degree is also related to the ratio between superlattice (odd,odd,odd)
and sub-cell (even,even,even) peaks of the XRD pattern \citep{liu2003}.
The results based upon both structure refinement and peak intensities
of the XRD pattern suggest that the Cu for Y substitution
may reduces the long-range ordering in BYT crystalline structure,
as is shown in both Table \ref{tab:rietveld} and Figure \ref{fig:ordering}.

The high dielectric permittivity observed in non-ordered double perovskites
is understood in terms of randomic cation/valence long-range order
distribution \citep{king2010}, \citep{yanez2009}. In the Ba$_{2}$Y$_{1-x}$Cu$_{x}$TaO$_{6-y}$
system reported here the ordering degree decreasing affects
the distribution of both compensating oxygen vacancies and hole carriers
within the grains \citep{adams2002}. It also offers considerable
contributions for the dielectric relaxation \citep{yang2010,mahato2017,rai2016,banerjee2019functional}.

\begin{figure}[H]
\centering \subfloat[]{\includegraphics[width=0.7\textwidth]{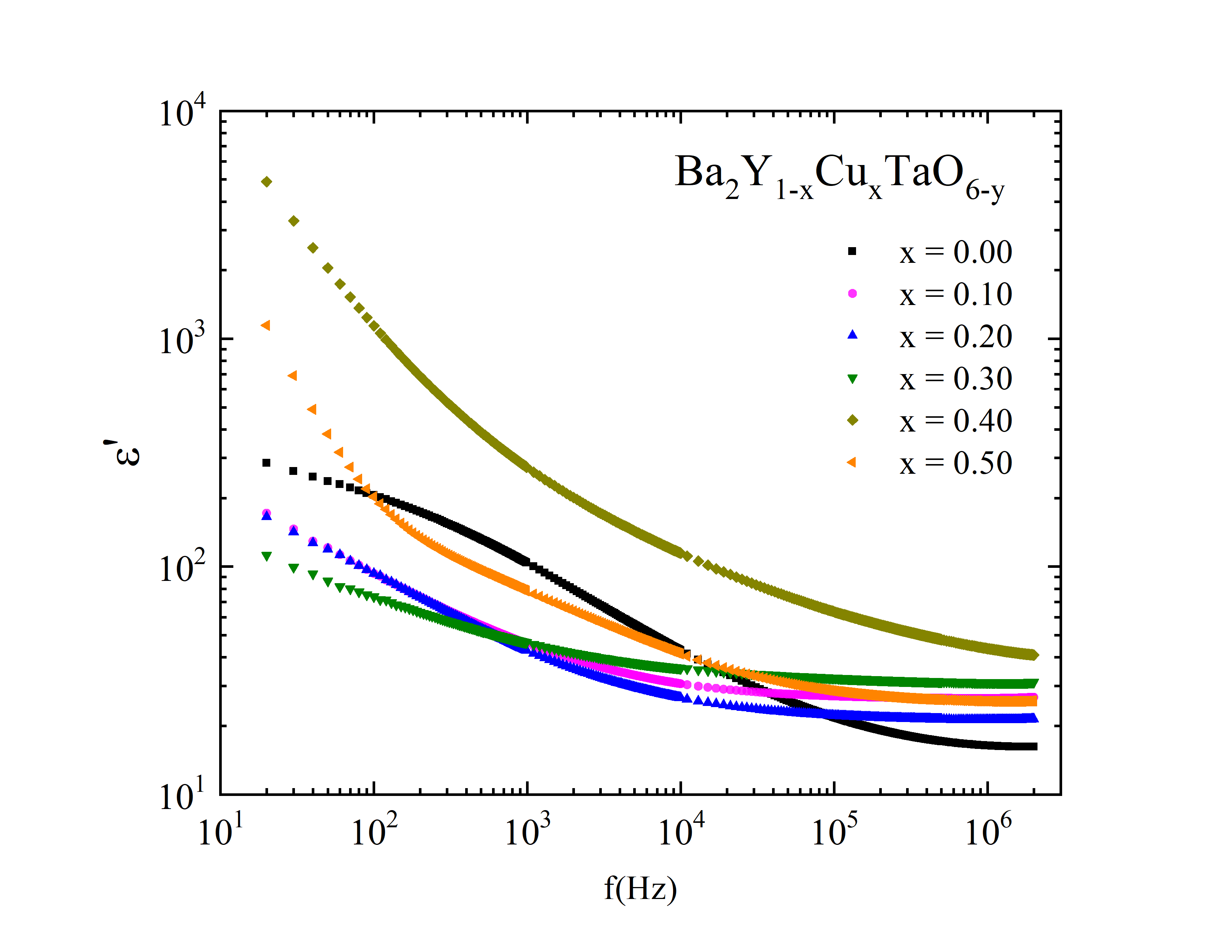}

}\\
 \subfloat[]{\includegraphics[width=0.7\textwidth]{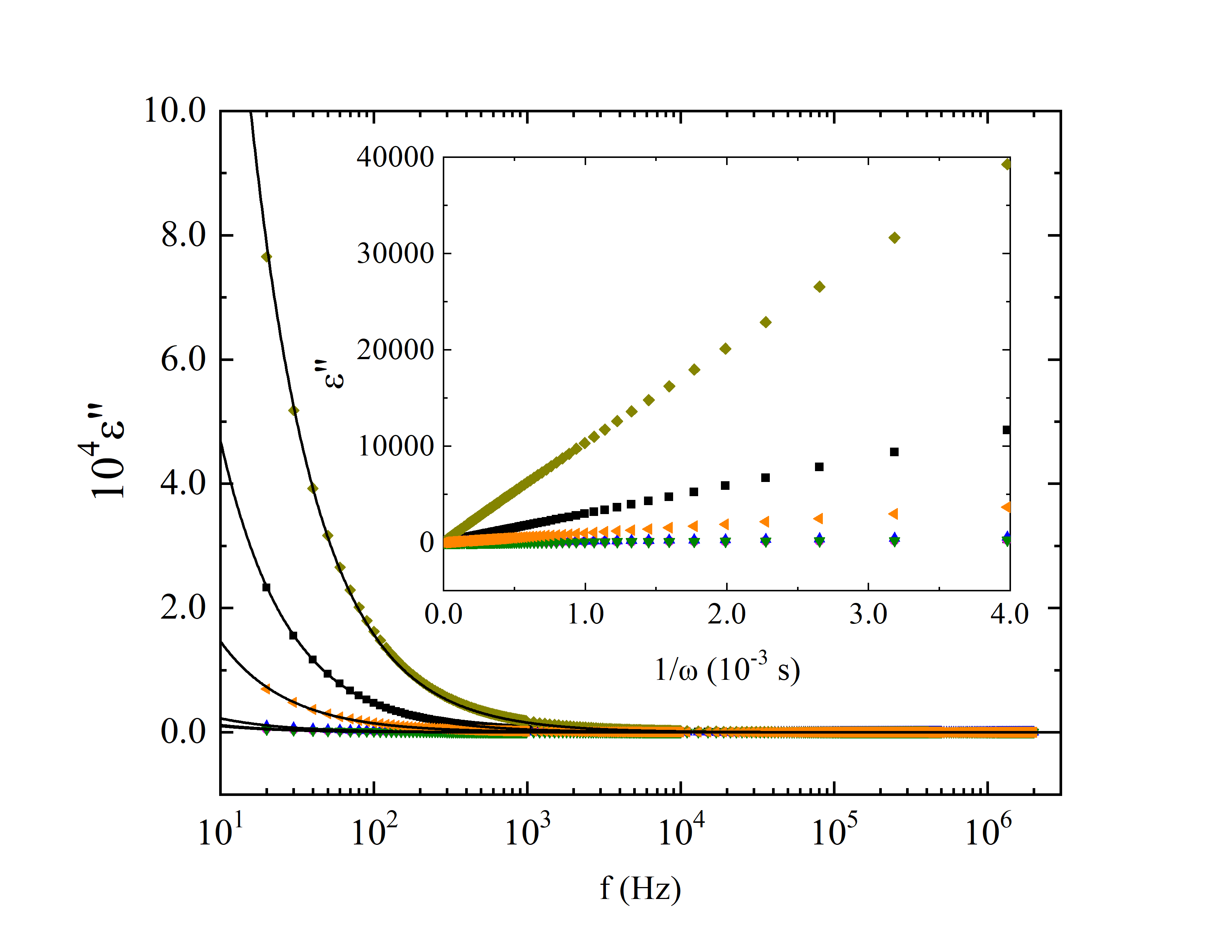}

}\caption{Variation of (a) Real (${\epsilon'}$) and (b) Imaginary (${\epsilon''}$)
parts of permittivity with the variation of frequency at room temperature,
the solid lines shows the fitting with Maxwell-Wagner model using
Equation \ref{eq:MW} and the inset shows ${\epsilon''}$ as function
as ${1/\omega}$.}
\label{fig:permittivity} 
\end{figure}

Figure \ref{fig:permittivity}(a) shows the frequency dependence of
real part of permittivity ($\epsilon'$) for Ba$_{2}$Y$_{1-x}$Cu$_{x}$TaO$_{6-y}$.
It is evident that the low frequency dielectric permittivity value
for x = 0.40 ceramics at room temperature is higher than for x = 0.00.
The initial high value of the real part of the dielectric permittivity
could be due to the drop of applied voltage across the thin grain
boundary widths and space charge polarization is generated in x =
0.40 and x = 0.50 samples, which enhances the dielectric constant
at the lower frequency region.

The initial high value of dielectric constant in Cu-doped samples
indicates the presence of dc conductivity due to Maxwell-Wagner (M-W)
relaxation process \citep{o2000}, which can be better visualized
in the imaginary part ($\epsilon''$) of the dielectric permittivity
shown in Figure \ref{fig:permittivity}b. In accordance with Maxwell-Wagner
(M-W) model, the imaginary part of dielectric permittivity can be
written with the following expression \citep{banerjee2017},

\begin{equation}
\epsilon"=\frac{1}{\omega C_{0}\left(R_{1}+R_{2}\right)}+\Delta\epsilon'\frac{\omega\tau}{1+\omega^{2}\tau^{2}},\label{eq:MW}
\end{equation}

where $\sigma=1/C_{0}\left(R_{1}+R_{2}\right)$ term is known as Ohmic
conductivity ($\sigma$), where $C_{0}$ is a geometric factor, and
$R_{1}$ and $R_{2}$ are the resistances of the real and imaginary
dielectric components, respectively \citep{banerjee2017,o2000}. The
magnitude of the Ohmic conductivity can be determined from the slope
of $\epsilon"$ vs. $1/\omega$ graph, as is shown in the Figure \ref{fig:permittivity}(b)
inset where one can see that x = 0.40 has the higher conductivity.
The eq. \ref{eq:MW} fits well with the experimental data which indicates
the presence of M-W polarization. Hence, the oxygen
vacancies generated due to the doping with Cu replacing Y
increased the ohmic conductivity for x=0.40 and x=0.50 samples as
they diffused into the grain boundary regions, which enhances their
dielectric constants at lower frequencies.

\begin{figure}[H]
\centering \subfloat[]{\includegraphics[width=0.7\textwidth]{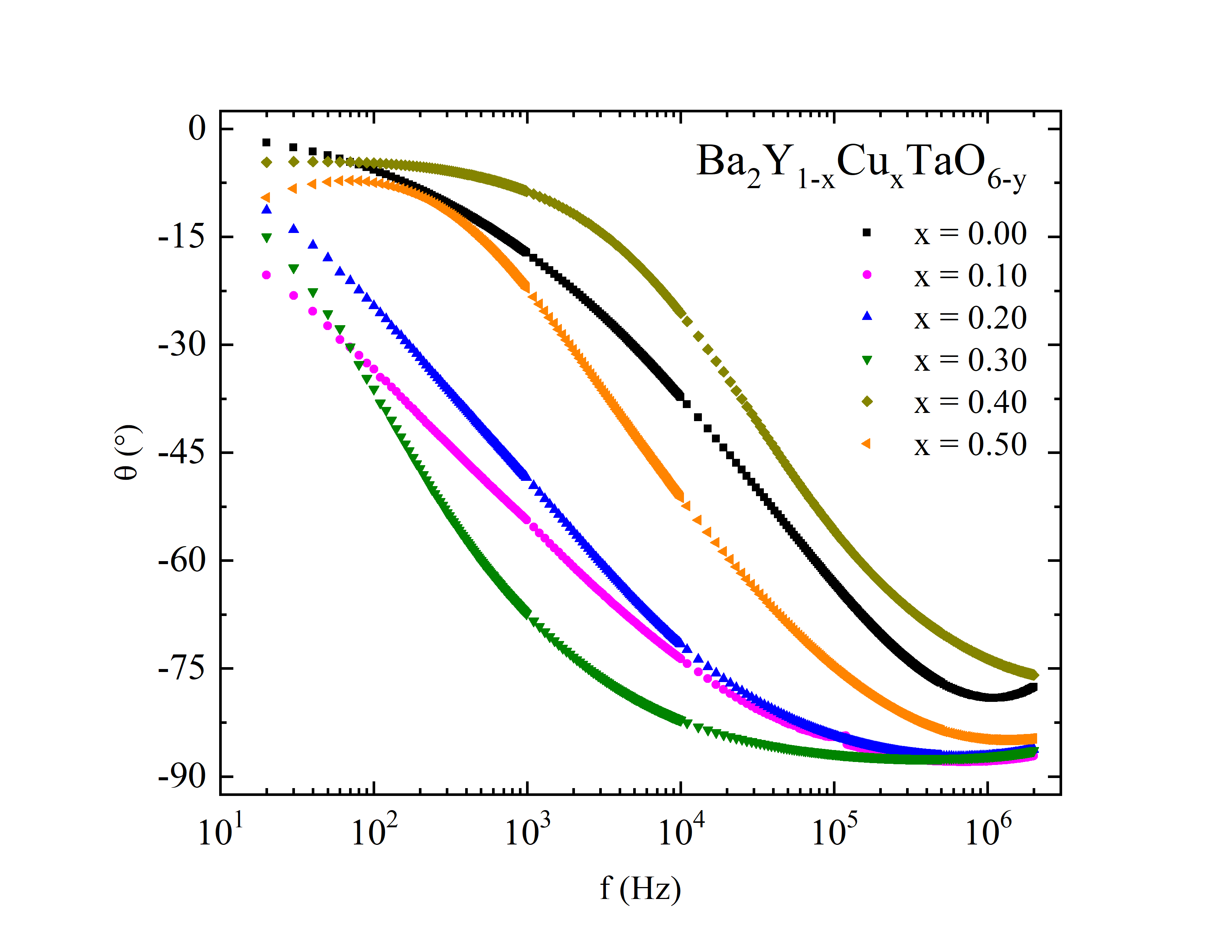}

}\\
 \subfloat[]{\includegraphics[width=0.7\textwidth]{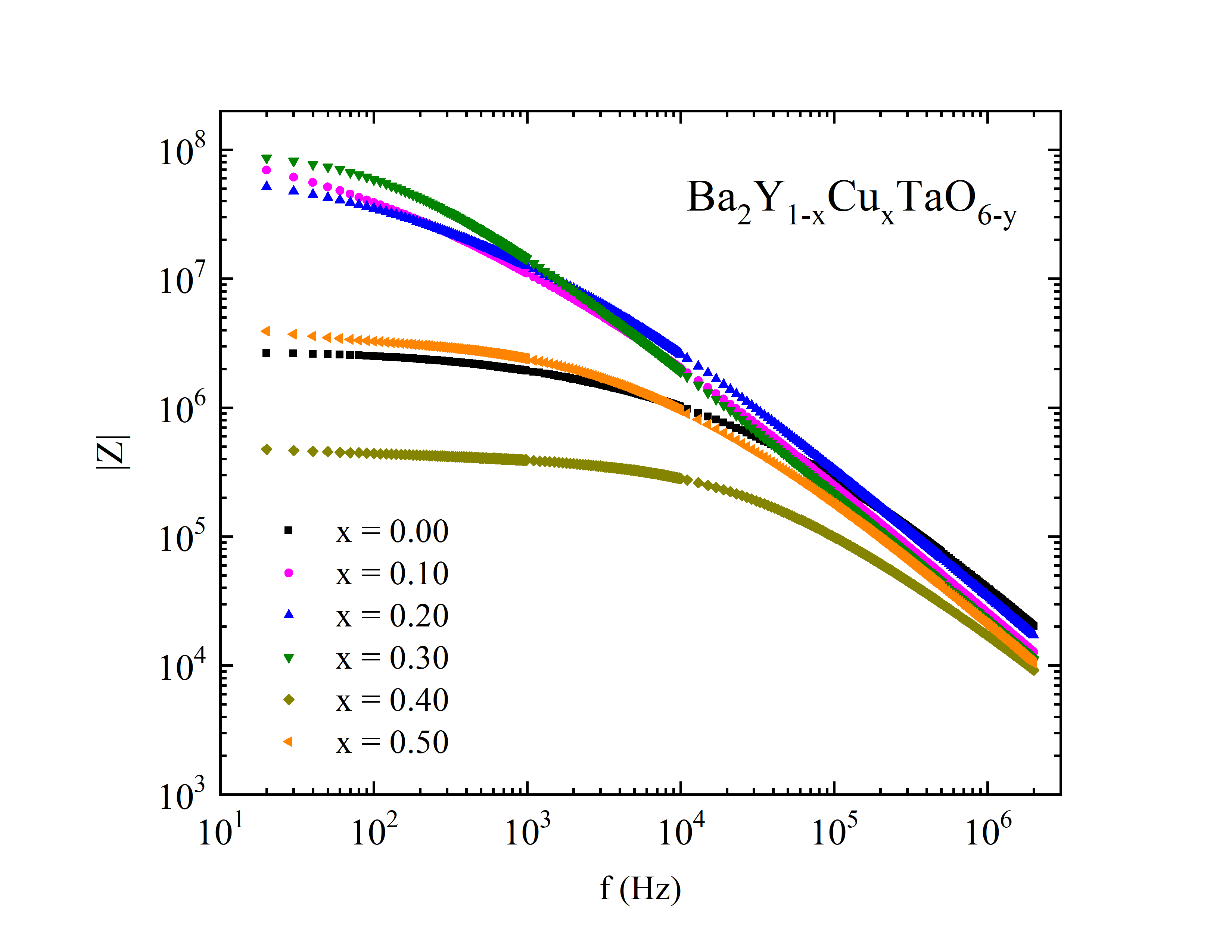}

}\caption{(a) Dependence of phase angle with frequency at room temperature and
(b) Dependence of real part of impedance (Z') with frequency at room
temperature.}
\label{fig:Z} 
\end{figure}

To understand the contribution of the interfacial polarization, complex
impedance spectroscopy (CIS) was performed at room temperature. The
phase angle of samples is shown in figure \ref{fig:Z}(a).

In general, the approach of phase angle towards 90$^{\circ}$ represents
the ideal poling state \citep{izquierdo2014}. So it can be observed
that slight addition of Cu enhances the poling condition
$\sim$ 87$^{\circ}$ in the x = 0.10 ceramic samples. But the magnitude
of phase angle for x = 0.40 was found to be $\sim$ 76$^{\circ}$. It suggests
that sufficient amount of Cu in the Y site of  Ba$_{2}$YTaO$_{6}$ crystalline
structure leads to changes in poling state and domain switching.

The dependence of the impedance with
frequency is shown in figure \ref{fig:Z}(b) on a double logarithmic
scale. It can be observed that the magnitude of Z' decreases gradually
for x = 0.10, x = 0.20, and x = 0.30 ceramics with the increase of
ac frequency \citep{kumari2015}. But for x = 0.00, x = 0.40, and
x = 0.50 ceramics the magnitude of $\vert Z \vert$ decreases gradually after 10\,kHz
frequency. The decrease of the real part of impedance at higher frequency
domain and thereafter gradual merger suggests a possible release
of space charge \citep{sutar2014} from the ceramics.

\begin{figure}[H]
\centering 

\subfloat[]{\includegraphics[width=1\textwidth]{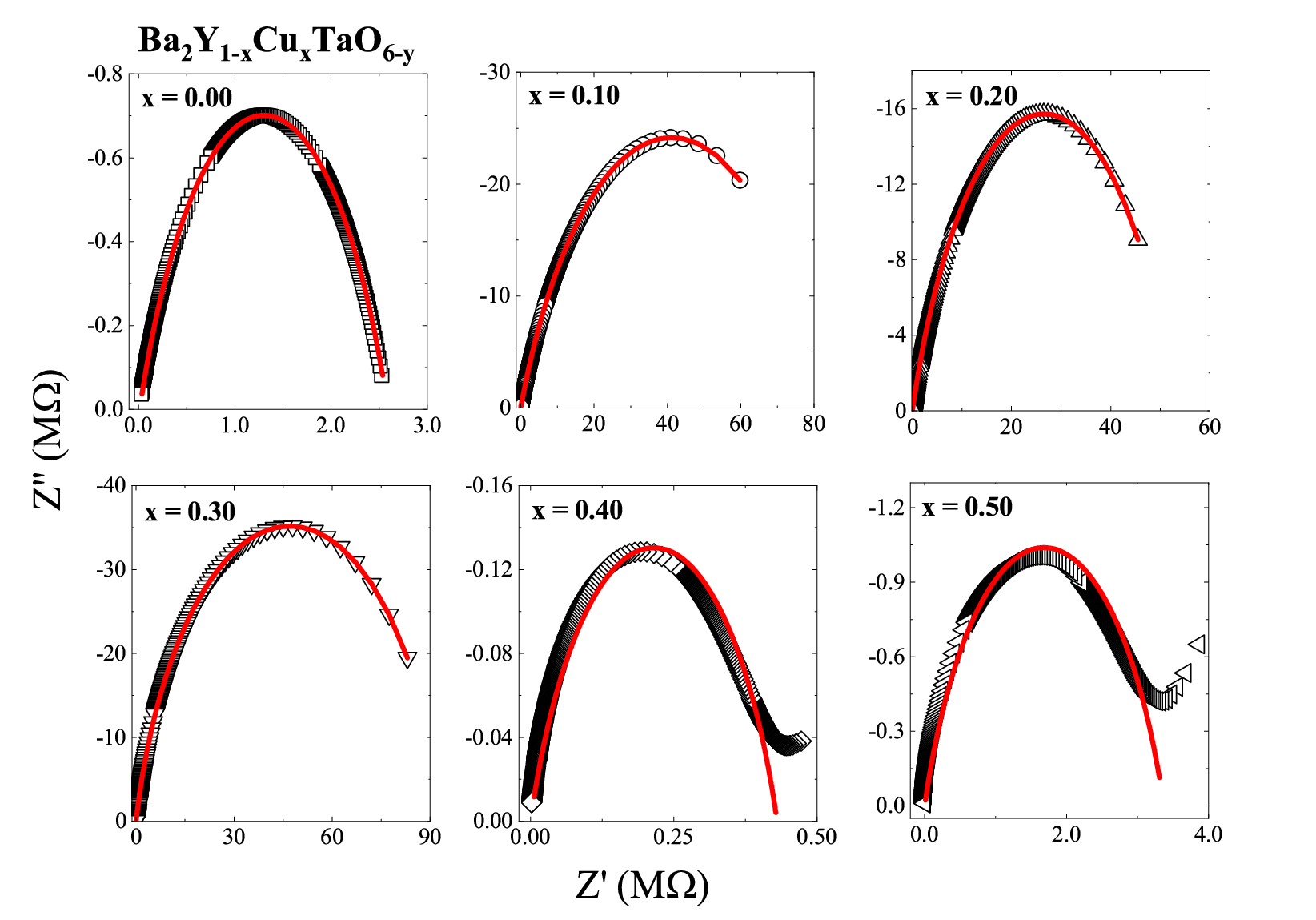}

} 

\subfloat[]{

\includegraphics[width=0.35\textwidth]{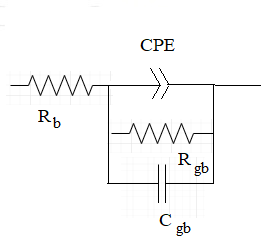}

}\caption{(a) Nyquist plots for Ba$_{2}$Y$_{1-x}$Cu$_{x}$TaO$_{6-y}$ ceramics
with $0.00\protect\leq x\protect\leq0.50$ (markers) and solid lines
are fits using Equation \ref{eq:Nyquist}. (b) Equivalent circuit used
for fitting. }
\label{fig:Nyquist} 
\end{figure}

Figure \ref{fig:Nyquist}(a) shows the Nyquist plots for Ba$_{2}$Y$_{1-x}$Cu$_{x}$TaO$_{6-y}$
ceramics. Since the observed semi circles are non-centered, Non-Debye
type relaxation i.e. Maxwell-Wagner relaxation exists in these ceramics
due to M-W relaxation \citep{barsukov2012}.

An equivalent circuit shown in figure \ref{fig:Nyquist}(b) may be represented
by a bulk resistance (R\textsubscript{b}) in series with a grain
boundary resistance (R\textsubscript{gb} ), grain boundary capacitance
(C\textsubscript{gb} ), and a constant phase element impedance (Z\textsubscript{CPE}
) in parallel as: 

\begin{equation}
Z(\omega)=R_{b}+\left[\frac{1}{R_{gb}}+\frac{1}{\frac{1}{C_{gb}(j\omega)}}+\frac{1}{\frac{1}{P(j\omega)^{n}}}\right]^{-1}\label{eq:Nyquist}
\end{equation}

In, $Z_{CPE}=1/P(j\omega)^{n}$ , P is the CPE parameter and n is
the CPE element which behaves like a double layer capacitor. The CPE
is identical to a capacitance when n=1 and to a simple resistance
when n=0 \citep{yu2019effect}. The experimental data fitted well
with Equation \ref{eq:Nyquist}, establishing the validity of the
equivalent circuit model. The single semicircle for each type of composition
indicates the single conductivity mechanism in the ceramics. Table
\ref{tab:electrical} shows the values of the fitting parameters for
the ceramics samples.

It can be observed from the magnitude of Table \ref{tab:electrical}
that all samples the CPE behaving like a parallel capacitor-resistor
in the equivalent circuit \citep{mahato2017}. With the addition of
Cu in Ba$_{2}$Y$_{1-x}$Cu$_{x}$TaO$_{6-y}$
both resistance and capacitance of grain boundaries increase gradually
which contribute to the barrier to the motion of charge carriers within
large domain bulks of electrical resistance orders of magnitude lower
than the boundaries resistance. Then, we conclude that it builds up
a space charge polarization across the boundary regions which was
represented by M-W model.

\section{Conclusions}

Ceramics samples of Ba$_{2}$Y$_{1-x}$Cu$_{x}$TaO$_{6-y}$ with $0.00\leq x\leq0.50$ were studied by X ray diffratometry, electronic scanning microscopy, dielectric permittivity measurements, and complex impedance spectroscopy. SEM images show liquid phase and defects induced by Copper. The Rietveld refinement of the XRD patterns reveal systematic changes in the crystalline structure, ordering degree, and domain sizes with the Cu content. The complex dielectric permittivity measurement demonstrated that the dielectric relaxation of  Ba$_{2}$Y$_{1-x}$Cu$_{x}$TaO$_{6-y}$ ceramics is described by the Maxwell-Wagner model. The complex impedance spectroscopy suggests that sufficient Cu for Y substitution in Ba$_{2}$YTaO$_{6}$ ceramics leads to changes in poling state and domain switching. The study also confirmed that the $x$ value ($0.00\leq x\leq0.50$) in Ba$_{2}$Y$_{1-x}$Cu$_{x}$TaO$_{6-y}$ ceramics affected resistance and capacitance of grain boundaries which
contributed to the barrier in motion of charge and build up a space charge polarization across the boundary regions.

\section*{Acknowledgments}

F. S. Oliveira acknowledges the Coordenação de Aperfeiçoamento de Pessoal de Nível Superior (CAPES) - finance Code 001. P. Banerjee acknowledges UGC, India for grant no. F.30-457/2018 (BSR).


\section*{Compliance with Ethical Standards}

Conflict of Interest: The authors declare that they have no conflict of interest.

\bibliographystyle{unsrt}
\bibliography{refs_BYTC}

\newpage{}

\newpage{}

\begin{table*}
\centering \caption{Rietveld refinement results as function as x in Ba$_{2}$Y$_{1-x}$Cu$_{x}$TaO$_{6-y}$.
Refinement agreement factors: $R_{exp}$ (expected) $R_{p}$ (profile),
$R_{wp}$ (pondered profile), and goodness of fit ($\chi^{2}$). Lattice
parameter (a),
crystallite size ($D$),
strain ($\epsilon$), ordering degree ($\eta$), and the ratio between
(111) and (220) peak intensities.}
\label{tab:rietveld} %
\begin{tabular}{p{2 cm}p{2 cm}p{2cm}p{2cm}p{2cm} p{2cm}p{2cm}}
\hline 
\multicolumn{7}{c}{Ba$_{2}$Y$_{1-x}$Cu$_{x}$TaO$_{6-y}$}\tabularnewline
\hline 
x  & 0.00  & 0.10  & 0.20  & 0.30  & 0.40  & 0.50 \tabularnewline
$R_{exp}\,(\%)$  & 5.11  & 5.69  & 5.13  & 6.48  & 4.62  & 5.13 \tabularnewline
$R_{p}\,(\%)$  & 8.58  & 9.01  & 9.45  & 6.13  & 4.31  & 7.3\tabularnewline
$R_{wp}\,(\%)$  & 17.55  & 15.96  & 18.36  & 7.42  & 5.63  & 14\tabularnewline
$\chi^{2}$  & 3.42  & 2.81  & 3.58  & 1.14  & 1.22  & 2.73 \tabularnewline
a\,({Å})  & 8.423(1)  & 8.433(1)  & 8.403(1)  & 8.379(1)  & 8.343(1)  & 8.342(3) \tabularnewline
D\,({Å})  & 973(229) & 830(96)  & 1041(236)  & 1031(201)  & 4338(2590)  & 4795(2723) \tabularnewline
$\epsilon\,(\%)$  & 0.05(4)  & 0.09(4)  & 0.04(2)  & 0.04(2)  & 0.09(3)  & 0.11(2) \tabularnewline
$\eta\,(\%)$  & 99.9  & 99.4  & 96.7  & 96.1  & 97.2  & 95.5 \tabularnewline
$I_{(111)}:I_{(220)}$  & 0.132  & 0.176  & 0.113  & 0.119  & 0.095  & 0.090 \tabularnewline
\hline 
\end{tabular}
\end{table*}

\begin{table*}
\centering \caption{Electrical parameters of the equivalent electrical circuit obtained
from complex impedance spectrum fits using Equation \ref{eq:Nyquist}
for BaY$_{1-x}$Cu$_{x}$TaO$_{6-y}$ samples}
\label{tab:electrical} 
\begin{tabular}{p{2cm}p{2cm}p{2cm}p{2cm}p{2.2cm}p{2cm}}
\hline 
x  & R$_{b}$\,($k\Omega$)  & R$_{gb}$\,(M$\Omega$)  & C$_{gb}$\,($pF$)  & P $\left(nFs^{n-1}\right)$  & n \tabularnewline
\hline 
0.00  & 4.06 & 2.82 & 3.157 & 2.81 & 0.484\tabularnewline
0.10  & 263.5  & 92.21 &  3.18 & 0.378 & 0.587\tabularnewline
0.20  & 63.5  & 58.76 & 2.743 & 0.363  & 0.571\tabularnewline
0.30  & 71.76  & 10.22  & 6.01 & 0.12  & 0.642\tabularnewline
0.40  & 4.74  & 0.47 & 9.672  & 12.2 & 0.435\tabularnewline
0.50  & 12.85  &  3.75 & 7.22  & 2.264 & 0.491\tabularnewline
\hline 
\end{tabular}
\end{table*}

\end{document}